\documentclass[12pt,preprint]{aastex}
\begin{document}

\title{The Outburst of the Blazar AO 0235+164 in 2006 December: Shock-in-Jet Interpretation}
\author{V.A. Hagen-Thorn \altaffilmark{1,4}, V.M. Larionov \altaffilmark{1,2,4}, 
S.G. Jorstad \altaffilmark{1,3}, A.A. Arkharov \altaffilmark{1,2},  
E.I. Hagen-Thorn \altaffilmark{1,2}, N.V. Efimova \altaffilmark{1,2}, 
L.V. Larionova \altaffilmark{1},  A.P. Marscher \altaffilmark{3}}

\altaffiltext{1}{Astronomical Institute of St. Petersburg State University, Universitetskiy Pr. 28, Petrodvorets, 198504, St. Petersburg, Russia; HTH-home@yandex.ru}
\altaffiltext{2}{Main (Pulkovo) Astronomical Observatory of RAS, Pulkovskoe Sh. 60, 196140, St. Petersburg, Russia; vlar@astro.spbu.ru}
\altaffiltext{3}{Institute for Astrophysical Research, Boston University,
725 Commonwealth Ave., Boston, MA 02215-1401; jorstad@bu.edu}
\altaffiltext{4}{Isaac Newton Institute of Chile, St.-Petersburg Branch}
\shorttitle{Outburst of A0235+164 in 2006 December}
\shortauthors{Hagen-Thorn et al.}
\begin{abstract}
We present the results of polarimetric ($R$ band) and multicolor photometric 
($BVRIJHK$) observations of the blazar AO 0235+16 during an outburst in 2006 
December. The data reveal a short timescale of variability (several hours), 
which increases from optical to near-IR wavelengths; even shorter variations are 
detected in polarization. The flux density correlates 
with the degree of polarization, and at maximum degree of polarization 
the electric vector tends to align with the parsec-scale jet direction. 
We find that a variable component with 
a steady power-law spectral energy distribution and very high optical 
polarization (30-50\%) 
is responsible for the variability. We interpret these properties of the blazar within
a model of a transverse shock propagating down the jet. In this case a small 
change in the viewing angle of the jet, by $\lesssim 1^o$, and a decrease
in the shocked plasma compression by a factor of $\sim$1.5 are sufficient to 
account for the variability.
\end{abstract}

\section{Introduction}
Blazars are active galactic nuclei (AGNs) that
possess extreme properties such as violent variability, high polarization,
relativistic jets, and bright $\gamma$-ray emission.
It is commonly thought that the radio through
optical emission in blazars is synchrotron radiation
that originates in the jets. Polarimetric observations provide 
crucial data for investigating the magnetic field structure
that forms the framework of such radiation. However, 
the polarization behavior is usually too complex for straightforward
interpretation. For example, long-term observations of BL~Lac
reveal a change of degree of polarization from 0.5\%
to $>$20\% with all possible electric vector position angles 
(EVPAs), although the distribution of EVPAs indicates the existence of a
preferential direction aligned with the parsec-scale jet
\citep{HT94}. Difficulties are even more severe for 
physical models aimed to explain both the brightness and polarization 
behavior. A connection between polarization and brightness is rarely seen
at optical wavelengths  \citep{SMITH96}.
However, there are strong indications that the optical
polarized emission is linked to the mm-wave VLBI core 
of blazars jets \citep{LS00,GRSS06,FRANI07,J07}.

 At radio wavelengths some progress in interpetation of both polarization and brightness variability
has been achieved by using a model in which a shock propagates down the jet 
\citep[e.g., ][]{HAA89}. At optical wavelengths, where polarization and brightness
can vary on timescale of a few hours \citep[e.g., ][]{HT72,MN96,CARINI06},
a proper test of models requires very intensive monitoring of both
polarization and brightness. Such simultaneous photometric 
and polarimetric measurements are difficult to
obtain, but they provide a vital tool to distinguish
among different models responsible for the variability.

We have observed the BL~Lac object AO~0235+164 (z=0.94) during the relatively short
outburst in 2006 December as part of the polarimetric and photometric
monitoring program of blazars that has been carried out at Saint Petersburg 
State University for many years. AO~0235+164 shows extreme blazar properties, such
as intraday variability \citep[e.g., ][]{R05} and very high apparent speeds in 
the radio jet, $\beta_{\rm app}=(26\pm 7)c$~ \citep{P06}, that have drawn 
the attention of many investigators to this object \citep{R01,R05,R06,JUNK04,HT07}. 
In addition, AO~0235+164 is one of the most highly polarized BL~Lac objects known. 
During the outburst in 1983 January the optical polarization of the object 
varied from 12\% to 24\% over a 24 hour period \citep{SMITH96}.  
Our polarimetric  monitoring program has caught the source in a similar 
state but with an even higher degree of polarization, $\gtrsim~30\% $. 
The polarimetric observations are enhanced by multicolor photometric 
measurements. Here we present the observations and show that the correlations 
that we have uncovered provide strong evidence in support of the shock model.  

\section{Observational data}

Observational data at optical wavelengths ($BVRI$ bands) were obtained at 
the 70-cm reflector of the Crimean Astrophysical Observatory with the 
photometer-polarimeter of the Astronomical Institute at St. Petersburg
State University based on an ST-7 CCD. Methods of observation and 
data reduction of the photometric mode are described in detail in \citet{HT06}.
We used the comparison stars recommended by the Whole Earth
Blazar Telescope (WEBT) program \footnote{http://www.to.astro/blazars/webt/}. 
The photometric errors do not exceed $0.03^m$ in $B$ band  and $0.02^m$
in other bands when the object is brighter than $17^m$.

Observations at near-IR wavelengths ($JHK$ bands) were carried out at the 
1.1-m telescope of the Main (Pulkovo) Astronomical Observatory of the Russian
Academy of Sciences located at Campo Imperatore (Italy) and equipped 
with the SWIRCAM camera. Methods of reduction
are the same as for the optical observations. Errors  in $JHK$ do not exceed
$0.02^m$ when the object is brighter than $16^m$ in J band. 
In all cases the calibration of \citet{MEAD90} was used
to transform magnitudes to flux densities
(subsequently referred to as ``fluxes''). 

Polarimetric observations were performed in $R$ band using two Savart plates
rotated by $45^\circ$ relative to each another. By swapping the plates,
the observer can obtain either the relative Stokes $q$ or $u$ parameter from the 
two split images of each source in the field.
Instrumental polarization was found via stars located near the object under
the assumption that their radiation is unpolarized. The Galactic latitude 
of AO~0235+164 is $-39^\circ$ so that interstellar polarization (ISP) in this direction is less
than 0.7\%. To correct for the ISP, the mean relative Stokes parameters  
of nearby stars were subtracted from the relative Stokes parameters of 
the object. This accounts for the instrumental polarization as well. 
The errors in degree of polarization are less than 2\% for
$R<16.5^m$, while the EVPA is determined with a precision of
several degrees.

\begin{figure}
\epsscale{.8}
\plotone{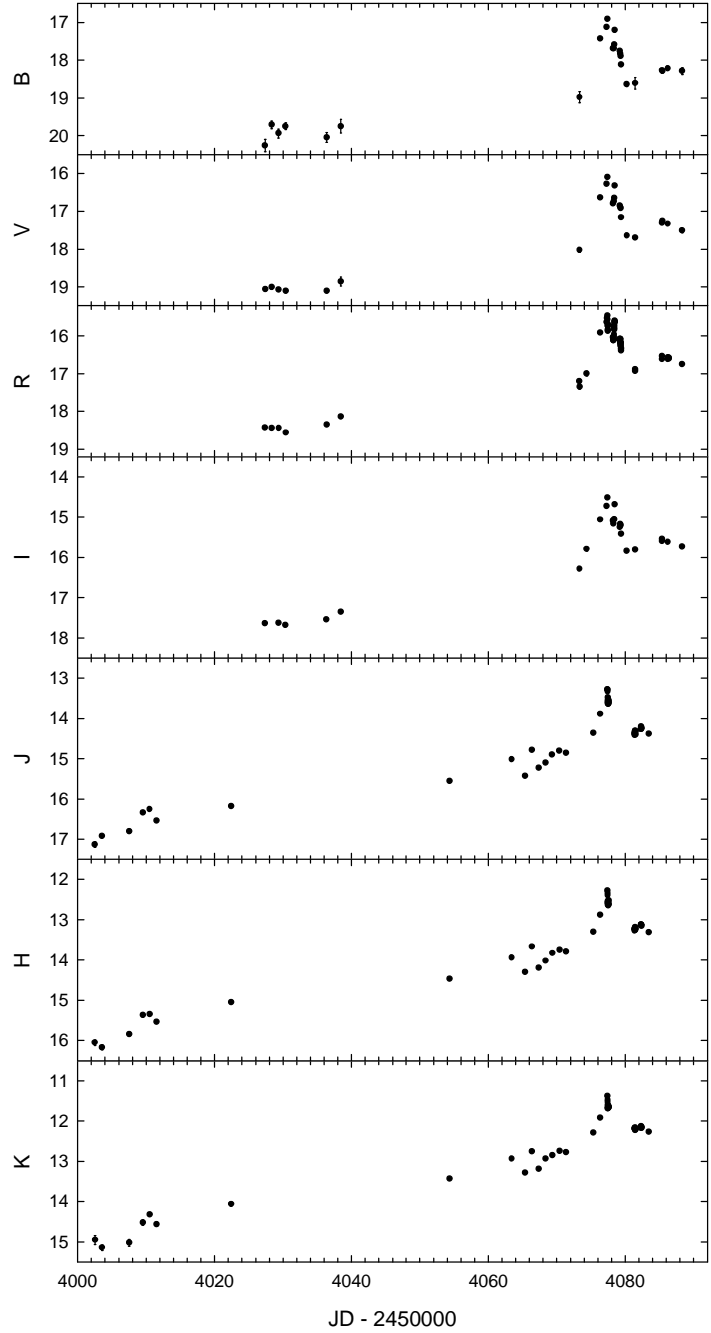}
\caption{Light curves of AO~0235+16 in the optical and IR bands.} 
\label{FDATA}
\end{figure}

\begin{figure}
\epsscale{0.8}
\plotone{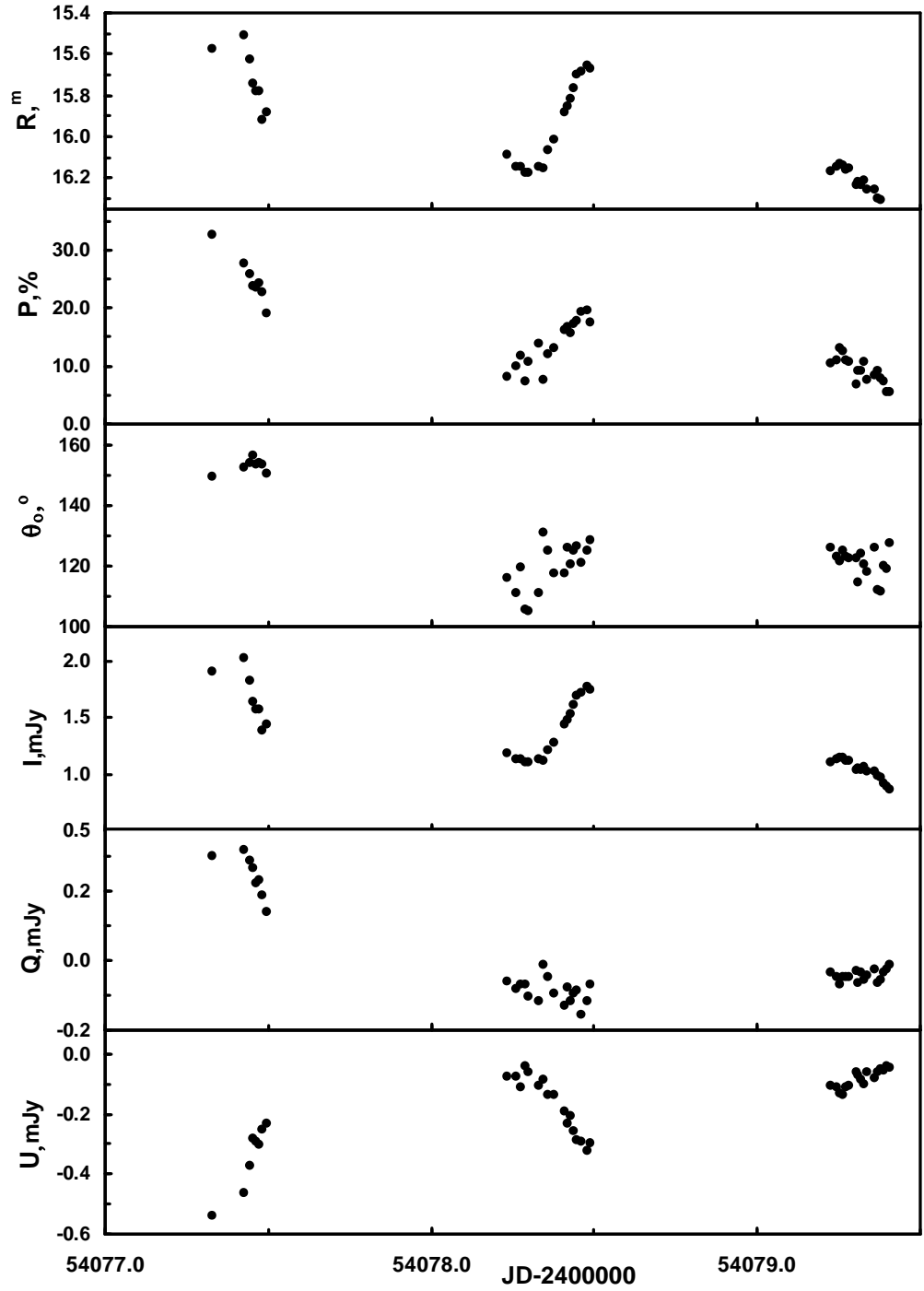}
\caption{Simultaneous photometric and polarimetric data for AO 0235+16 at three
consecutive Julian dates during the outburst.} 
\label{PDATA}
\end{figure}

The results of photometric observations in the optical and IR bands are given in
Tables \ref{TAB_1} and \ref{TAB_2}, respectively. The light curves are shown in Figure 
\ref{FDATA} for the convenience of visual inspection.
One can see that during the outburst the
brightness increases by more than $3^m$. Table \ref{TAB_3} 
lists the results of the polarimetric observations. 
In Figure \ref{PDATA} the upper panels illustrate the photometric $R$ and
polarimetric observations for three consecutive Julian dates 
during the most dramatic period of the outburst. Intraday variations (IDV) are 
evident within a given Julian date. The bottom panels
present absolute Stokes parameters during the same period. 
The existence of systematic variations during each night 
is a prominent feature of all the curves.

\section{Global Properties of Variability}
\subsection{Timescale of Variability}
We have determined the timescale of variability of the flux in each band
using the definition proposed by 
\citet{BJO74}: $\tau=dt/{\rm ln(F_1/F_2)}$, where $dt$ is the time interval
between flux measurements $F_1$ and $F_2$, with $F_1>F_2$. We have calculated 
all possible timescales $\tau_{\rm ij}$ for any pair of
observations for which $|F_i-F_j|>\sigma_{F_i}+\sigma_{F_j}$ at frequency $\nu$.
The minimum timescale at frequency $\nu$ is determined as 
$\tau_\nu={\rm min}\{\tau_{\rm ij,\nu}\}$, where $i$=1,...$N-1$, $j$=i+1,...$N$, 
and $N$ is the number of observations. Table \ref{TAB_var}
lists the results of these calculations as well as the timescale of the polarized flux 
variability. The uncertainties of the values of $\tau_\nu$ are determined 
by errors of the flux measurements. However, the minimum timescale of variability
that we derive is mainly constrained by the sampling, 
for example, the $B$, $V$, and $I$ light curves have 
only two measurements during JD~2454077 (Table \ref{TAB_1}), the date at which
we find the minimum timescales in $R$ band and at IR wavelengths. In Table \ref{TAB_var}
we enclose the timescales derived from these light curves in brackets
to indicate that they can not be compared with the timescales derived from
the better sampled $R$, $J$, $H$, and $K$ light curves. 

The values of $\tau_\nu$ indicate that during the event the timescale 
of variability was only a few hours and the shortest timescale, $\sim$0.3~hr, 
occured in the polarized flux. Comparison of the timescales of the total
and polarized flux variability in $R$ band implies that the polarized emission
originates in a subsection ($\sim$1/5) of the optical emission region.
Moreover, the light curves with similar sampling during the event 
($R,J,H$, and $K$ data) reveal an increase of timescale with wavelength.

\subsection{Multicolor Behavior}
\begin{figure}
\epsscale{1.0}
\plotone{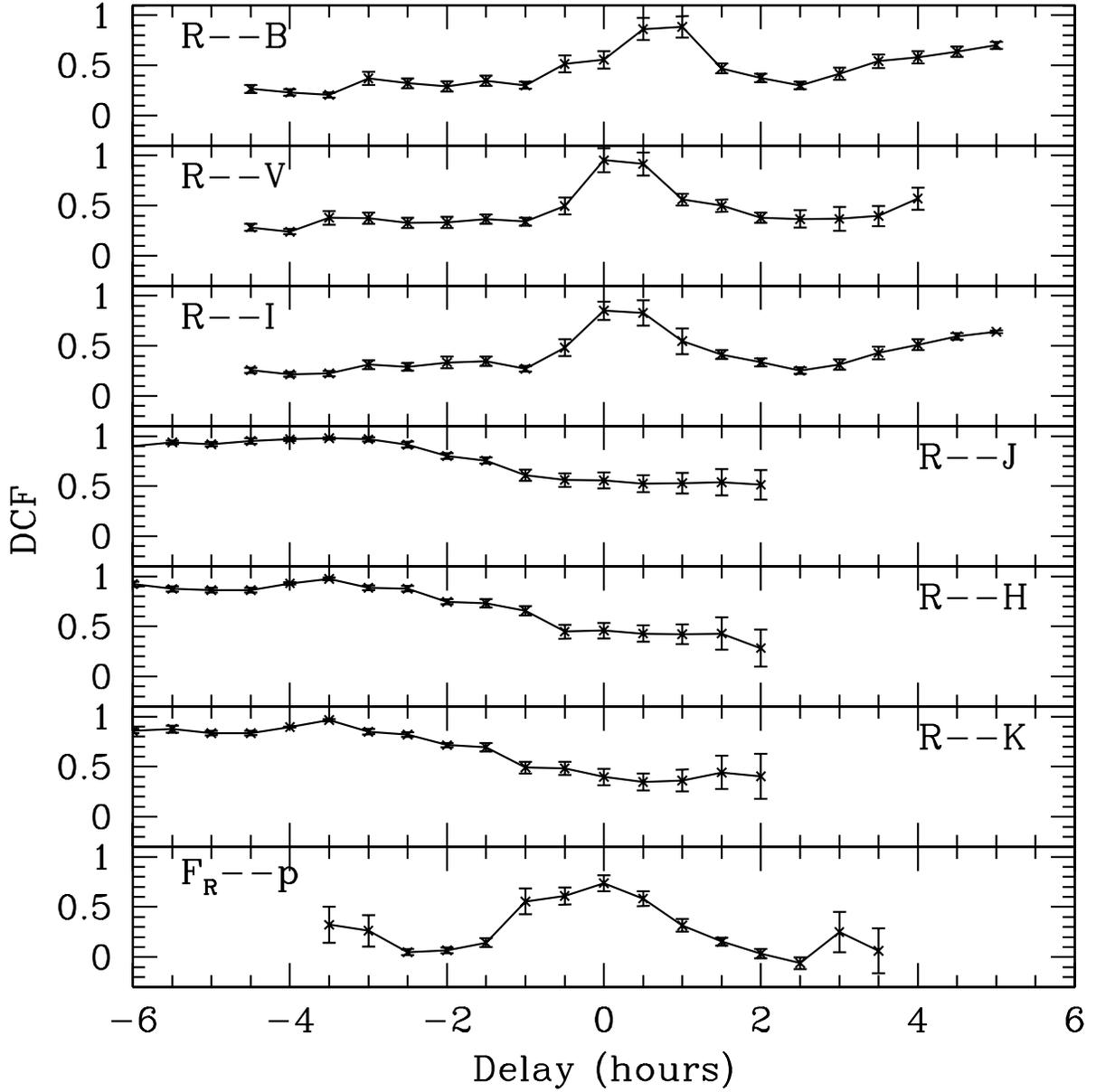}
\caption{Discrete cross-correlation function between light curves
in different bands with respect to the $R$-band light curve; a negative delay
means that $R$ band variations lead. {\it Bottom panel}: DCF between degree
of polarization and total flux density in $R$ band.
} 
\label{cor}
\end{figure}

We have calculated the discrete cross-correlation function, $DCF$, 
between the $R$ band light curve and those at other wavelengths  using the 
approach suggested by \citet{EK98} for unequally spaced data.
The functions were computed for lags between light curves from
$-6$ to $+6$~hr with a step of $0.5$~hr and with the constraint
that at least 5 pairs of measurements are needed to define the coefficient
of correlation at a given delay. Figure \ref{cor} presents the results 
of the calculations. The absence of points at some lags indicates that 
the data are not capable of determining the $DCF$ for such delays 
owing to the constraint.

Figure \ref{cor} shows that the $DCF$s have maxima at different
delays between the light curves. The maxima are distinct only
for the optical light curves, and indicate that variations in $B$ band
lead variations in $R$ band by $\sim$0.5-1~hr, while variations
in $V$ and $I$ bands are simultaneous within 0.5~hr with the variability
in $R$ band. Although the peaks of the $DCF$s for the near-IR light
curves are broad and their positions are uncertain within 4-5~hrs,
the delay of variations in the near-IR region compared to those
in $R$ band is statistically significant. The peaks of the $DCF$ and their delays 
for each light curve with respect to the $R$ band light curve are
listed in Table \ref{TAB_var}. Such delays between the light curves
are expected if the sizes of the emission regions at different wavelengths
scale with the timescales of the variability found above.

\subsection{Polarization Properties}
\begin{figure}
\epsscale{1.0}
\plotone{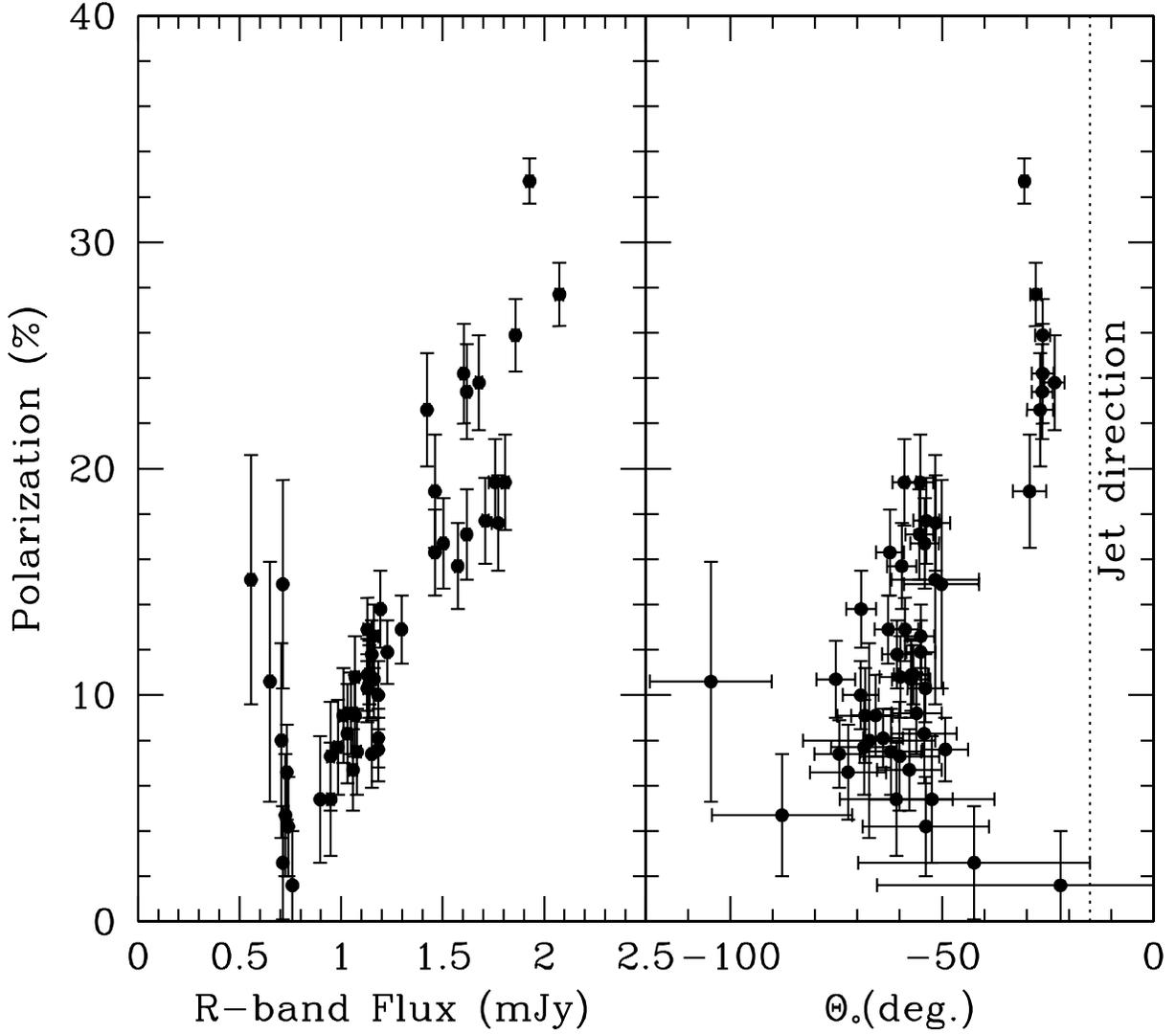}
\caption{{\it Left panel}: Dependence between the degree of polarization and flux 
in $R$ band. {\it Right panel}: Dependence between the degree of 
polarization and polarization position angle; the dotted line indicates 
direction of the parsec scale jet.} 
\label{PTheta}
\end{figure}

Table 3 shows that very high degrees of polarization  (up to 30\%) were observed 
during the December 2006 outburst. Figure \ref{PTheta} demonstrates  that  the degree of 
polarization increased when the object brightened. The latter is confirmed 
by the high value 
of the $DCF$ peak between the flux in $R$ band and the degree of polarization
(Table \ref{TAB_var}). The $DCF$ indicates also that the $R$ band and fractional polarization
variations were simultaneous within 0.5~hr (Fig. \ref{cor}).
Figure \ref{PTheta} shows a wide range, $\sim$90$^\circ$, of variability of 
the polarization position angle. However, $\Theta_\circ$ aligns closely with 
the parsec jet direction, $\sim -15^\circ$ \citep{J01,P06}, when the 
degree of polarization exceeds 20\%. 

Visual inspection of Figure \ref{PDATA} reveals a correlation between Stokes
parameters - between $Q$ and $I$ and between $U$ and $I$ -- and different behavior
of the Stokes parameters from one epoch to another.
The correlation coefficients calculated separately 
for each Julian date, given in Table \ref{TAB_Pvar},  
confirm the tight correlation between the Stokes parameters.

\section{Properties of the Source Responsible for Variability}
\subsection{Method of Analysis}
Our current understanding of active galactic nuclei pictures an AGN as a complex system in 
which different components - the host galaxy, accretion disk, corona above the disk, broad line 
regions, and jet - contribute to the total flux. In some cases multicolor photometric and polarimetric 
observations of variability may make it possible to derive the characteristics of 
the variable source (spectral energy distribution, SED, and polarization 
parameters) without knowledge of the absolute contribution of the other
components to the total radiation. 

Let us suppose that the variability within some time interval is due to a single variable component. 
If the variability is caused only by its flux variation but the relative Stokes parameters $q$ and $u$ 
(for polarimetry) or relative SED (for photometry) remain unchanged, 
then in the space of absolute Stokes parameters $\{I, Q, U\}$ or in the $n$-dimensional flux space $\{F_1, ..., F_n\}$ ($n$ is the number of
spectral bands used in multicolor observations) the observational points must lie on straight lines. 
The slopes of these lines are relative Stokes parameters of the variable component (for polarimetry) 
or flux ratios for different pairs of bands as determined by its SED (for photometry).

With small caveats, the opposite is also true. In the case of photometry,
a linear relation between observed fluxes in two different wavelengths during some 
period of flux variability implies that the flux ratio does not change. 
Such a relation for several bands indicates that the relative SED of 
the variable component 
remains steady and can be derived from the slopes of the lines. In the case of polarimetry, 
a linear relation between observed Stokes parameters
suggests that the variable component has constant
degree of polarization and polarization position angle, the values of which
can be obtained from the relative Stokes parameters $q$ and $u$ that
correspond to the slope of the $Q$ vs. $I$ and $U$ vs. $I$ lines, respectively.
Further details of the method of analysis are provided in \citet{HM99}.
  
\subsection{Polarization Parameters of the Variable Component}

\begin{figure}
\epsscale{0.5}
\plotone{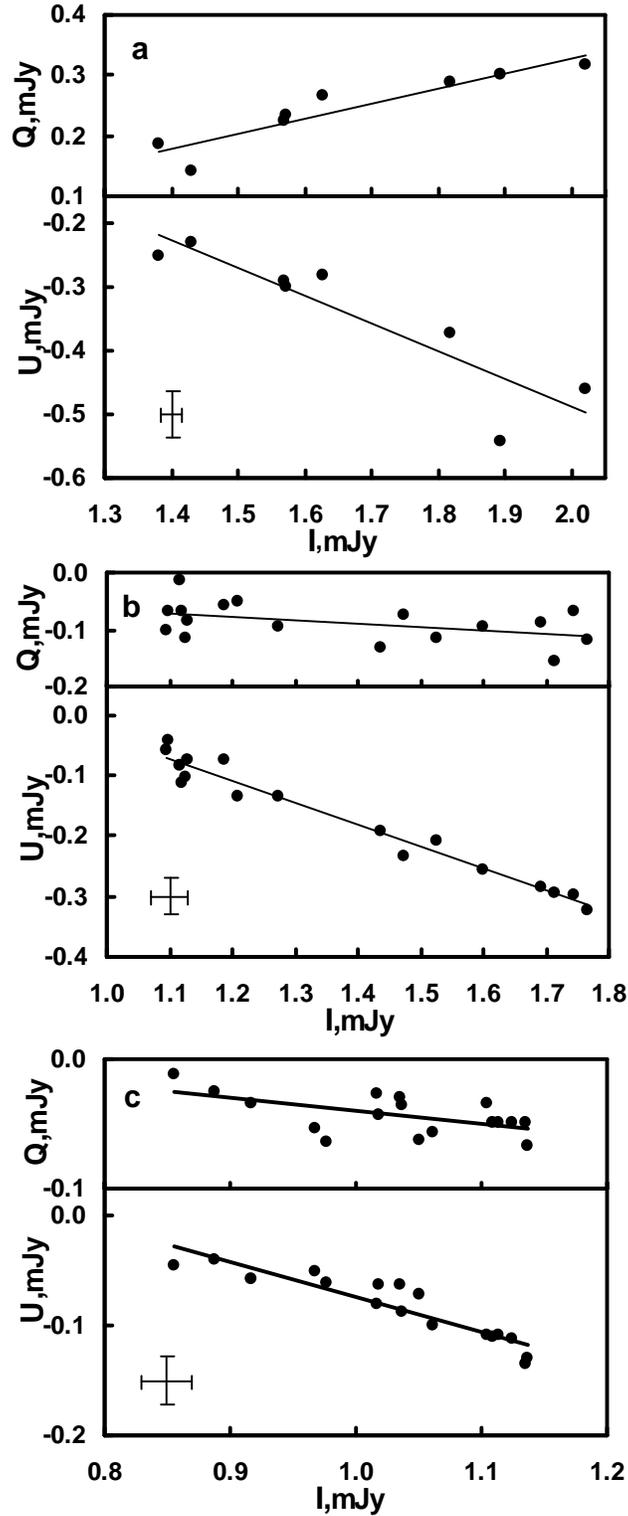}
\caption{Comparison of absolute Stokes parameters for three 
consecutive Julian dates during the outburst (a - 2454077, b - 2454078, c - 2454079); crosses correspond to the uncertainties of the parameters.}
\label{STOKES}
\end{figure}

Correlation plots of the absolute Stokes parameters for three consecutive epochs 
during the outburst (JD 2454077, 2454078, and 2454079) are given in Figure \ref{STOKES} (a,b, and c,
respectively). One can see that in all cases the points lie on straight lines that 
leads us to conclude that during a given epoch the relative Stokes parameters 
of the variable component remained unchanged. The slopes of the straight lines in Figure 
\ref{STOKES} are determined by the orthogonal regression method. They give the relative 
Stokes parameters $q_{var}$ and $u_{var}$ of the variable component that we use to calculate 
its degree and position angle of polarization, $p_{var}$ and 
$\Theta_{var}$ (Table \ref{TAB_Pvar}). Table \ref{TAB_Pvar} indicates 
that the degree of polarization of the variable component is very high and decreases 
from 50\% to 30\% as the object fades. This is accompaned by a systematic variation of 
the polarization direction. At the highest degree of polarization the EVPA tends 
to align to within 15$^\circ$ with the direction of the parsec-scale jet 
inferred from the VLBI maps of AO 0235+164 \citep{J01,P06}. 

\subsection{Spectral Energy Distribution of the Variable Component}
\begin{figure}
\epsscale{1.0}
\plotone{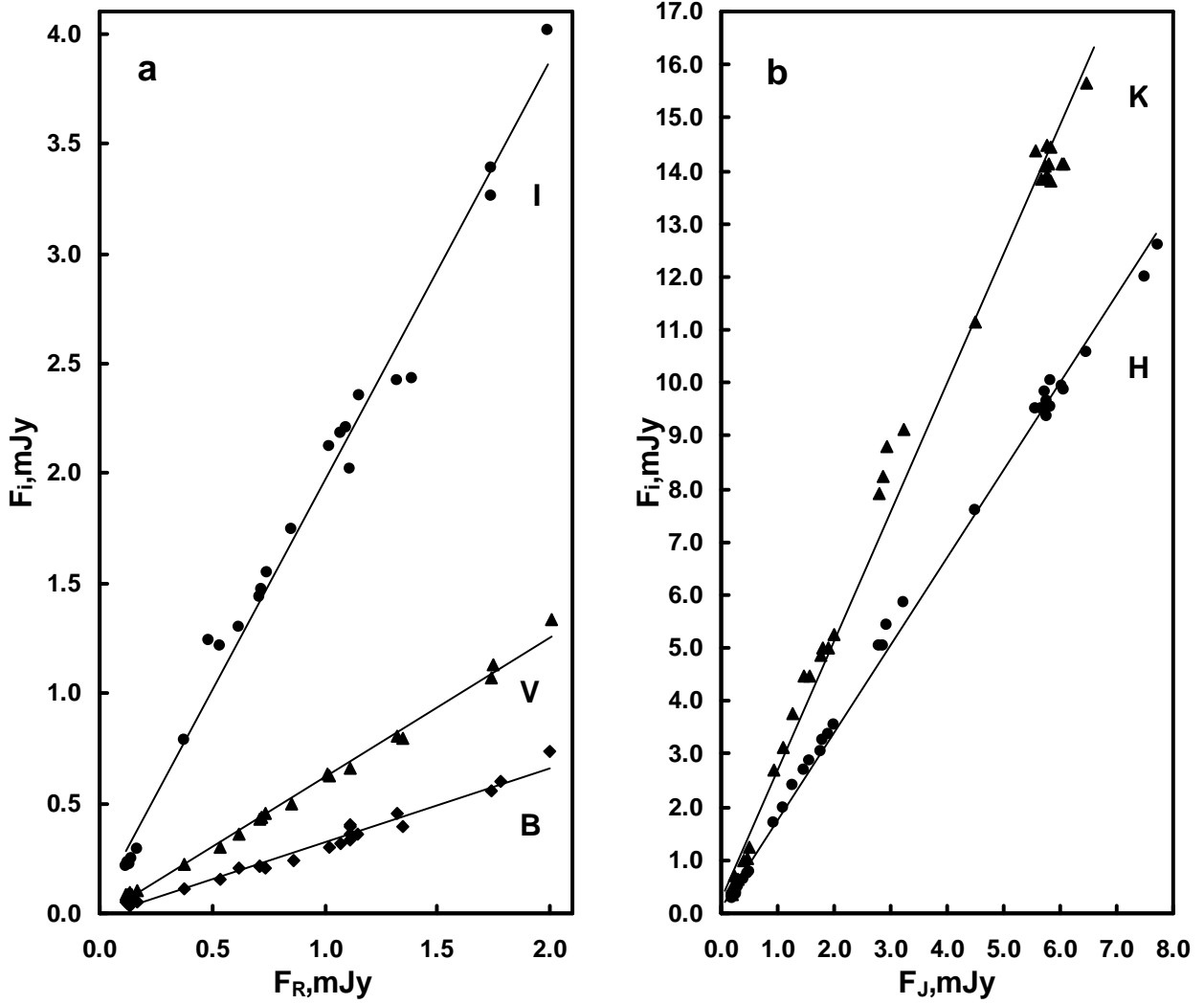}
\caption{Flux-flux diagrams for optical (a) and (b) IR wavelengths.} 
\label{Fcompar}
\end{figure}

We compare the fluxes in different bands relative to the $R$ and $J$ fluxes for 
the optical and near-IR data, respectively (see Fig. \ref{Fcompar}), since these 
two bands have 
the largest number of observations. According to Figure \ref{PDATA}, the data reveal 
intraday variability that can 
affect the flux-flux comparison because the measurements are not completely simultaneous. 
For example, during Julian dates with intensive monitoring, delays of $\sim10-15$
minutes between observations in different bands occur. 
For such epochs the fluxes in $R$ band were interpolated to 
the corresponding times of the $B$, $V$, and $I$ bands measurements 
(the same for $J$ band with respect to $H$ and $K$ bands). For epochs with a few measurements, 
the average at each band was used. 

\begin{figure}
\epsscale{1.0}
\plotone{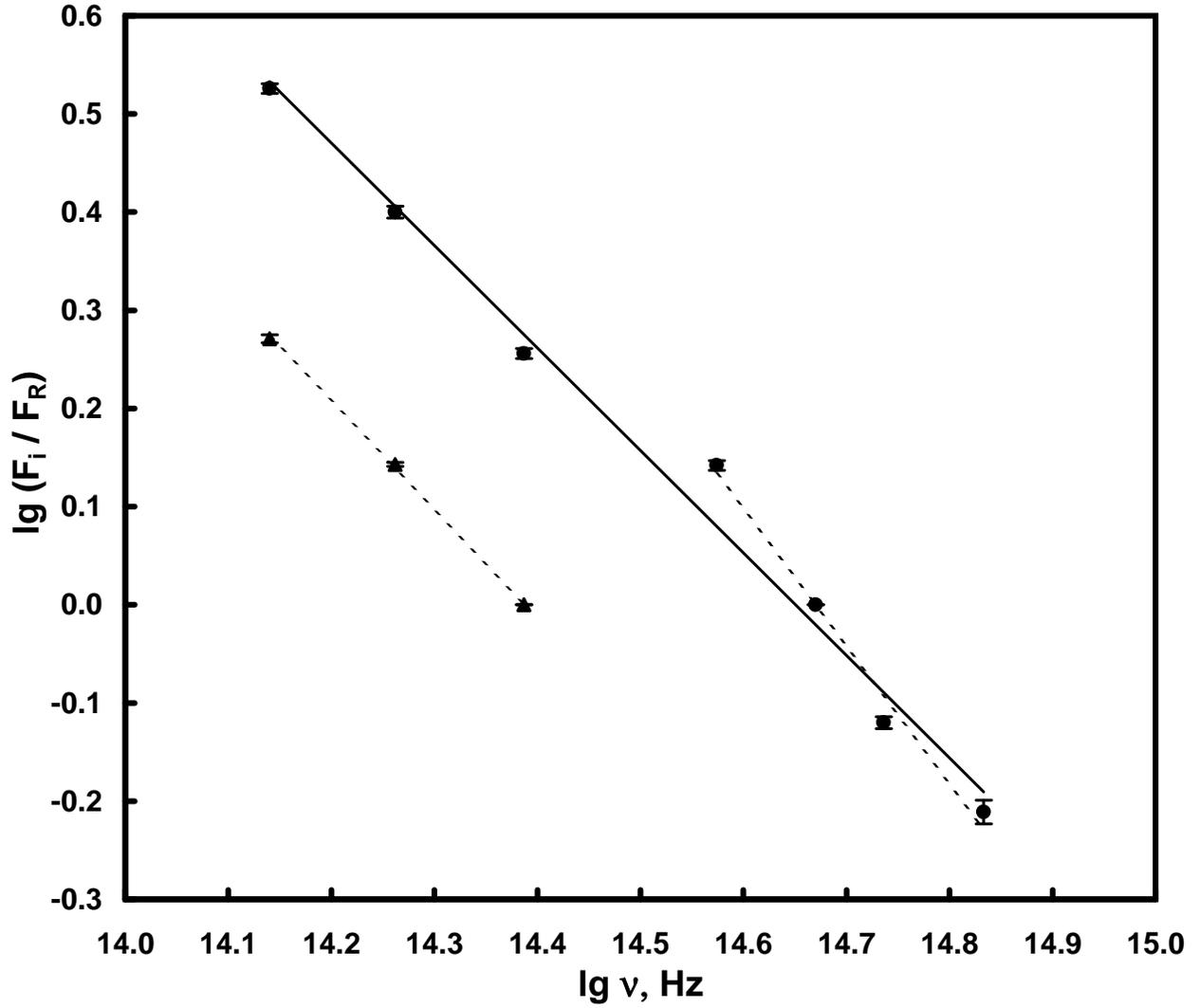}
\caption{The relative SEDs at IR and optical wavelengths (dashed lines) and 
for the combined spectrum (solid line). The optical SED is relative to the $R$ band
flux. The IR SED is relative to the $J$ band flux. The combined SED is 
constructed using a linear dependence between simultaneous observations in $J$ and $R$ bands.} 
\label{Spectr}
\end{figure}

The slopes of the lines were derived using the orthogonal regression method. They 
are corrected for extinction based on the information provided by \citet{R05}. 
The logarithms of the corrected values are listed in Table \ref{TAB_var}. 
Using these values, we have constructed SEDs
in the optical and near-IR regions as well as the combined SED 
(via the flux-flux diagram for $R$ and $J$) for the whole range from $B$ to $K$ bands (Fig. \ref{Spectr}).
Figure \ref{Spectr} shows that the spectra can be represented by a power law, 
$F_{\nu} \sim \nu^{-\alpha}$ , where $\alpha$ is the spectral index. In the near-IR region 
$\alpha = 1.10 \pm 0.03$, while in the optical region $\alpha = 1.39 \pm 0.12$. Although there appears to be a slight steepening of the spectrum in the optical region, this
falls  within the 2$\sigma$ 
uncertainty, so that the combined SED can be fit by a single power law with 
$\alpha = 1.04 \pm 0.06$. This indicates that the same component with the
relative SED shown in Figure \ref{Spectr} is responsible for
the source variability in the whole range from $B$ to $K$ bands.
     
\section{Discussion}
The results of the polarimetric and photometric observations shows that 
the dramatic variability 
seen in AO~0235+164 in 2006 December might be explained by a single variable component that 
possesses the following properties:
(i) steady spectral shape;
(ii) increase of size of the emitting region with wavelength;
(iii) very high degree of polarization, up to 50\%; 
(iv) strong correlation between degree of polarization and
flux; and (v) tendency of the magnetic field at the highest 
polarization state to be perpendicular to the inner radio jet direction. 
These properties are characteristics of the model that 
describes an outburst in a blazar as a transverse shock 
propagating down a relativistic jet with turbulent magnetic field 
\citep{MG85,HAA85,MAR96}. 
The timescale of variability relates to the thickness of 
the shock front, which is determined by the lifetime 
of relativistic electrons accelerated at the front. This thickness decreases 
with frequency owing to synchrotron radiation losses. The shock
orders the turbulent magnetic field along the front, which is
perpendicular to the jet direction.  

We consider a model of a square-wave transverse shock moving through a
turbulent plasma with a constant Lorentz factor, $\Gamma$.
The size and shape of the emission region at a given frequency are steady.
In the observer's frame the flux of the shocked region is boosted as
follows:
\begin{equation}
  F=F_\circ\nu^{-\alpha}\delta'^{(2+\alpha)}\delta^{(3+\alpha)}, \label{eq1}
\end{equation}
where $\delta'$ is the Doppler factor of the shocked
plasma in the rest frame of the shock front and $\delta$ is the Doppler
factor of the shock front in the observer's frame.
Determination of $\delta'$ requires knowledge of the speed and equation of
state of the plasma behind the shock front \citep{CAW91}, which is difficult
to specify due to a lack of the necessary information. Instead, we assume
that the velocity of the shocked plasma in the frame of the shock front $\ll c$ and
$\delta'\approx$1. We can then approximate that
$\delta=[\Gamma(1-\beta~cos\Phi)]^{-1}$, where
$\beta=\sqrt{1-\Gamma^{-2}}$ is the speed of the shock and $\Phi$ is the
angle between the line of sight and the velocity of the centroid
of the emission region.

The degree of polarization depends 
on the viewing angle of the shock, $\Psi$, and ratio of the density in the 
shocked region with respect to unshocked region, $\eta=n_{\rm shock}/n_{\rm unshock}$ \citep{HM91}:
\begin{equation}
p\approx \frac{\alpha+1}{\alpha+5/3}~\frac{(1-\eta^{-2})\sin^2\Psi}{2-(1-\eta^{-2})\sin^2\Psi}. \label{eq2}
\end{equation}
The viewing angle of the shock in the
observer's frame is subjected to relativistic 
aberration and determined by the bulk Lorentz factor 
and viewing angle of the jet:
\begin{equation}
\Psi=\tan^{-1}[\sin\Phi/(\Gamma(\cos\Phi-\sqrt{1-\Gamma^{-2}}))] \label{eq3}
\end{equation}

\begin{figure}
\epsscale{1.0}
\plotone{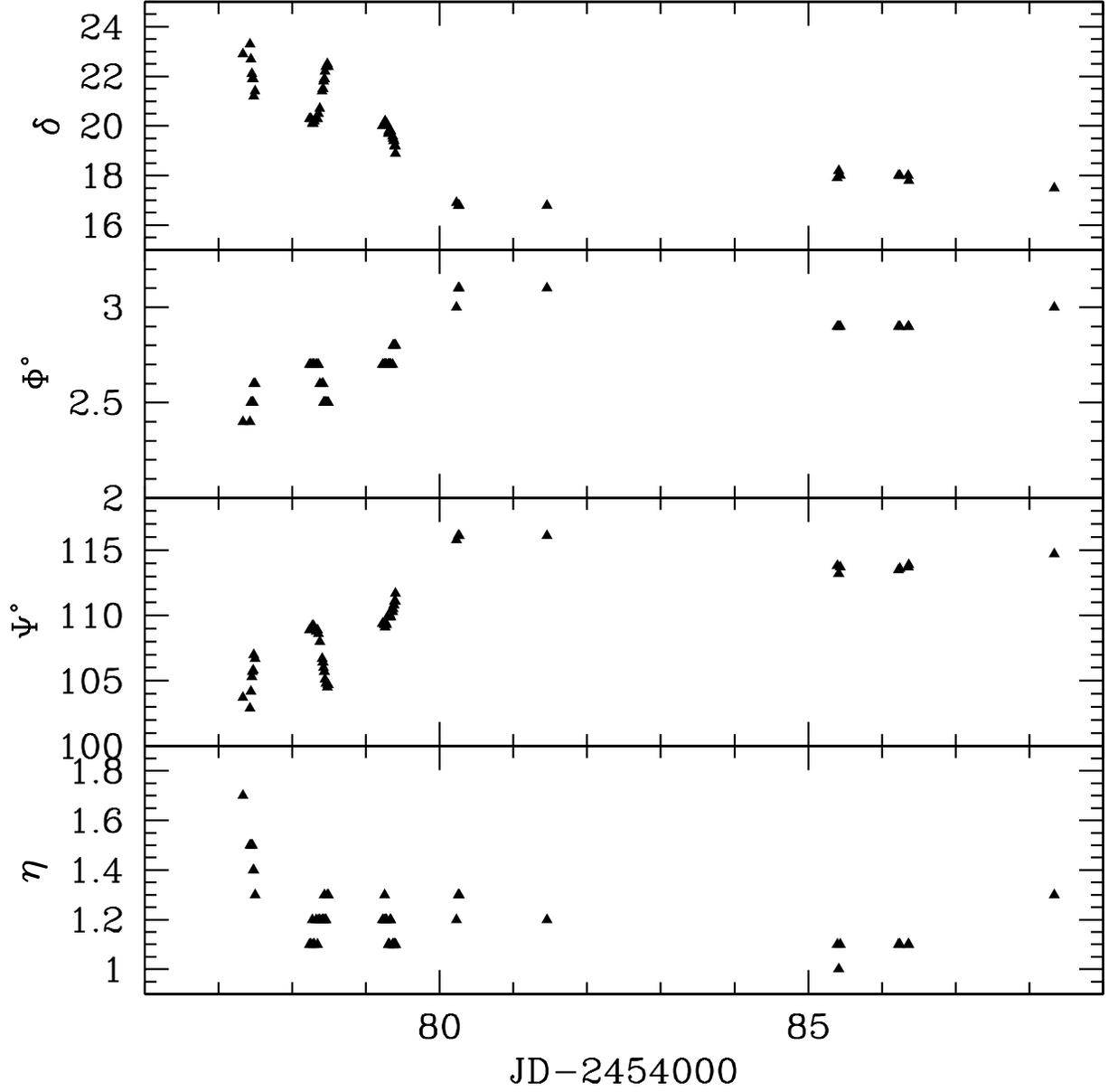}
\caption{The derived values of Doppler factor, $\delta$, angle between the jet
axis and line of sight, $\Phi$, viewing angle of the shock, $\Psi$, and compession
factor of the shocked plasma, $\eta$.} 
\label{model}
\end{figure}

We use the derived spectral index of the variable component $\alpha=1.04$  
and assume $\Gamma=30$, consistent with the highest reliable apparent speed seen 
in the jet of AO~0235+16 \citep{J01,P06}.  We apply equation (\ref{eq1}) 
to the photometric data in the $R$ band to estimate the Doppler factor
as a function of time. The scaling factor $F_\circ$ is determined under the assumption
that $F_\circ=F_{\rm max}\nu^{\alpha}/\delta_\circ^{(3+\alpha)}$, where
$F_{\rm max}$ is the maximum observed flux in $R$ band and $\delta_\circ$ is 
obtained for $\Phi_\circ$, which is calculated 
from equations (\ref{eq2}) and (\ref{eq3}) for the maximum derived
degree of polarization of the variable component (Table \ref{TAB_Pvar}). 
In this case we adopt $\eta$=2.5 that is the minimum possible compression 
that produces polarization as high as 50\%. This yields $\delta_\circ=23.3$
at the maximum polarization.
Figure \ref{model} shows the values of the Doppler factor and viewing angle
of the jet derived from the observed flux variability.   
The latter allows us to estimate the viewing angle of the shock (eq. \ref{eq3})
and value of the plasma compression as functions of time using the light curve 
of fractional polarization (eq. \ref{eq2}). The values of $\Psi$ and $\eta$ are 
shown in Figure \ref{model} as well.

Figure \ref{model} demonstrates that the observed flux and polarization
behavior can be explained by a small change in the viewing angle $\lesssim 1^\circ$
and by variations of shocked plasma compression while the bulk Lorentz
factor remains constant. A small viewing angle ($\sim 2.5^\circ$) agrees 
with the highly core-dominated
radio images of the blazar. A change in the viewing angle can be caused either
by curvature of the jet by $\lesssim 1^\circ$ or by a curved trajectory of the
shock-causing perturbation 
inside the jet for an opening angle $\gtrsim 1^\circ$. In the brightest 
flux state the viewing angle is the nearest to the line of sight, the shock
is seen almost edge-on ($\Psi\sim 103^\circ$), the plasma compression
is the highest ($\eta\sim 1.7$), and the parameters change on timescale
of hours. In the end of the outburst  $\eta\sim 1.1$ and the parameters
are steady within a week. Changes in the rate of injection of 
relativistic electrons could be related to variations in $\eta$.

In this scenario a shock is responsible for
the injection of the relativistic electrons into the emitting region,
leading to a steady spectral shape and a short timescale of 
the variability in the optical and IR bands.
Therefore, we assume that the total
flux and linear polarization variability is caused by changes in the
viewing angle and in injection of
relativistic electrons on timescales of hours. Radiative losses
determine the size of the emitting region behind the shock front,
which is of the order of 2-3 light hours (Table \ref{TAB_var}).
A steady spectral shape implies a quasi-steady state of the emission,
which is re-established every 2-3~hrs, between the rate of injection 
and the radiative losses. 
In the observer's frame the synchrotron lifetime of electrons emitting 
at frequency $\nu$ is:
\begin{equation}
t_{\rm loss}\approx 4.75\times 10^2{\left[\frac{1+z}{\delta\nu_{GHz}B_G^3}\right]}^{1/2}~~{\rm days}, \label{eq4}
\end{equation}
where $B_G$ is the magnetic field in Gauss. 
Since $t_{\rm loss}\approx\tau_\nu$,  for $\delta=23.3$ 
(the maximum derived Doppler factor) and $\tau_\nu=1.7$~hr in the $R$ band 
(Table \ref{TAB_var}), eq.(\ref{eq4}) yields an estimate of 
the magnetic field,  $B\approx 0.5~G$, 
that agrees with other estimates of the 
typical magnetic field in a blazar core
\citep[e.g., ][]{MG85}. 
Equation (\ref{eq4}) implies also that the timescale of variability
in $K$ band should be longer than that in $R$ band
by a factor of 1.8, consistent with the values of $\tau_\nu$ 
derived from the observations. 
The highest degree of polarization, $p_{\rm var}=50$\%,
is close to the maximum possible polarization (75\%)
for a synchrotron source
with $\alpha=1.04$ and a uniform magnetic field \citep{PAH70}.
This indicates an extremely well ordered  magnetic field
behind the shock front along the plane transverse to the jet direction.
Therefore, we consider that the polarimetric and multicolor photometric
behavior of AO~0235+164 during the outbust in 2006
December is consistent with the properties of the
shock-in-jet model. 

\section{Conclusion}
In 2006 December the BL~Lac object AO~0235+164 underwent a spectacular
outburst with an increase in optical flux by a factor of 4 and
with degree of polarization reaching 30\%. Analysis of the 
multicolor photometric data and polarization measurements in $R$
band indicate that:

(i) The flux increase has an achromatic character.

(ii) The spectral energy distribution of the variable component
is a power law with $\alpha$=1.04$\pm$0.06. 
 
(iii) Variations on timescales of hours are present at all wavelengths.
The minimum timescale of variability increases with wavelength. 

(iv) There is a delay between the variations at different wavelengths,
with fluctuations occurring first in $B$ band and last at the near-IR 
wavelengths.

(v) The fractional polarization correlates with the flux.

(vi) The direction of polarization at the highest polarization states tends
to align with the inner radio jet direction.

These properties of the variable component 
responsible for the outburst match the characteristics of a transverse
shock propagating down the jet. We find that the observed photometric
and polarimetric variability can be reproduced in detail
by the shock propagating down a twisted jet such that the viewing angle 
changes by $\lesssim 1^\circ$ and the plasma compression decreases
by a factor of $\sim$1.5 while the bulk Lorentz factor, $\Gamma=30$, remains constant.    

\acknowledgments
This work is supported by the Russian Fund of Basic Research, grant 05-02-17562.
S. Jorstad and A. Marscher acknowledge support from the 
National Science Foundation under grant no. AST-0406865.

\clearpage

\begin{deluxetable}{lrlrlrlr}
\singlespace
\tablecolumns{8}
\tablecaption{Optical Data \label{TAB_1}}
\tabletypesize{\footnotesize}
\tablehead{
\colhead{MJD}&\colhead{$B$}&\colhead{MJD}&\colhead{$V$}&
\colhead{MJD}&\colhead{$R$}&\colhead{MJD}&\colhead{$I$}\\
\colhead{(1)}&\colhead{(2)}&\colhead{(3)}&\colhead{(4)}&\colhead{(5)}
&\colhead{(6)}&\colhead{(7)}&\colhead{(8)}
}
\startdata
27.3865&20.26$\pm$0.17&27.3946&19.06$\pm$0.05&27.3784&18.43$\pm$0.03&27.3823&17.65$\pm$0.03 \\
28.3267&19.71$\pm$0.17&28.3345&19.01$\pm$0.07&28.3489&18.44$\pm$0.04&28.3473&17.63$\pm$0.03 \\
29.3514&19.95$\pm$0.17&29.3592&19.08$\pm$0.05&29.3749&18.45$\pm$0.04&29.3735&17.68$\pm$0.03 \\
\enddata
\tablecomments{MJD is JD-2454000; the table appears in full form electronically.}
\end{deluxetable}
\begin{deluxetable}{lrlrlr}
\singlespace
\tablecolumns{6}
\tablecaption{Near-IR Data \label{TAB_2}}
\tabletypesize{\footnotesize}
\tablehead{
\colhead{MJD}&\colhead{$J$}&\colhead{MJD}&\colhead{$H$}&
\colhead{MJD}&\colhead{$K$}\\
\colhead{(1)}&\colhead{(2)}&\colhead{(3)}&\colhead{(4)}&\colhead{(5)}
&\colhead{(6)}
}
\startdata
979.3865&17.24$\pm$0.11&979.5627&16.45$\pm$0.15&979.5666&15.31$\pm$0.19 \\
980.3267&17.25$\pm$0.07&980.6206&16.32$\pm$0.09&980.6258&15.47$\pm$0.16 \\
985.3514&17.27$\pm$0.06&985.5778&16.19$\pm$0.05&985.5828&15.70$\pm$0.16 \\
\enddata
\tablecomments{MJD is JD-2453000; the table appears in full form electronically.}
\end{deluxetable}
\begin{deluxetable}{lrr}
\singlespace
\tablecolumns{3}
\tablecaption{Polarization Data in $R$ band\label{TAB_3}}
\tabletypesize{\footnotesize}
\tablehead{
\colhead{MJD}&\colhead{$p$,\%}&\colhead{$\Theta^\circ$} \\
\colhead{(1)}&\colhead{(2)}&\colhead{(3)}
}
\startdata
77.3307&32.7$\pm$1.0&149.5$\pm$0.8 \\
77.4282&27.7$\pm$1.4&152.2$\pm$1.3 \\
77.4433&25.9$\pm$1.6&153.8$\pm$1.8 \\
\enddata
\tablecomments{Columns are as follows: (1) - JD-2454000;
(2) - degree of the polarization; (3) - position angle of the 
polarization; the table appears in full form electronically.}
\end{deluxetable}

\begin{deluxetable}{lrrrrrr}
\tablecolumns{7}
\tablecaption{Multicolor Properties of the Variable Component \label{TAB_var}}
\tabletypesize{\footnotesize}
\tablehead{
\colhead{Band}&\colhead{$lg(\nu)$}&\colhead{$N$}&\colhead{$\tau_{\rm min}$,hr}
&\colhead{$DCF_{\rm peak}$}&\colhead{$\Delta t$,hr}&\colhead{$lg(F_\nu/F_R)$} \\
\colhead{(1)}&\colhead{(2)}&\colhead{(3)}&\colhead{(4)}&\colhead{(5)}&\colhead{(6)}&
\colhead{(7)}
}
\startdata
$B$&14.833&24&(5.27$^{+1.36}_{-0.90}$)&0.89$\pm$0.10&$+1.0^{+0.5}_{-1}$&$-$0.211$\pm$0.012 \\
$V$&14.736&24&(6.15$^{+0.40}_{-0.34}$)&0.95$\pm$0.11&$0.0^{+1}_{-0.5}$&$-$0.120$\pm$0.006 \\
$R$&14.670&66&1.73$^{+0.22}_{-0.29}$&1.0&0.0&0.0              \\
$I$&14.574&24&(4.81$^{+0.29}_{-0.25}$)&0.85$\pm$0.09&$0.0^{+1}_{-0.5}$&0.142$\pm$0.007 \\
$J$&14.387&55&2.91$^{+0.46}_{-0.34}$&0.98$\pm$0.02&$-3.5^{+1.5}_{-2.5}$&0.256$\pm$0.007 \\
$H$&14.262&55&2.19$^{+0.55}_{-0.37}$&0.98$\pm$0.03&$-3.5^{+1.5}_{-2.5}$&0.400$\pm$0.006 \\
$K$&14.140&54&3.01$^{+1.21}_{-0.67}$&0.97$\pm$0.03&$-3.5^{+1.5}_{-2.5}$&0.526$\pm$0.005 \\
$P$&14.670&54&0.31$^{+0.04}_{-0.04}$&0.74$\pm$0.08&$0.0^{+1}_{-1.5}$&\nodata         \\
\enddata
\tablecomments{Columns are as follows: (1) - filter of photometry,
$P$ corresponds to the polarization data;
(2) - logarithm of effective frequency of the filter; 
(3) - number of observations;
(4) - minimum timescale of variability, unreliable values are enclosed
in brackets; (5) - peak of discrete cross-correlation 
function of the light curve at 
frequency $\nu$ with respect to $R$ band light curve; in the case 
of polarization data the peak is given for the $DCF$ between the $R$-band and
fractional polarization light curves; (6) - delay of the peak; 
$+$ means that the light curve at frequency $\nu$
leads the $R$ band light curve, while $-$ is the opposite case; (7) - logarithm of
the contribution of the variable component in the SED at frequency $\nu$
relative to the contribution in $R$ band.}
\end{deluxetable}

\begin{deluxetable}{lrrrr}
\singlespace
\tablecolumns{5}
\tablecaption{Polarization Properties of the Variable Component \label{TAB_Pvar}}
\tabletypesize{\footnotesize}
\tablehead{
\colhead{MJD}&\colhead{$r_{QI}$}&\colhead{$r_{UI}$}&\colhead{$p_{\rm var}$,\%}&\colhead{$\Theta_{\rm var},^\circ$} \\
\colhead{(1)}&\colhead{(2)}&\colhead{(3)}&\colhead{(4)}&\colhead{(5)} 
}
\startdata
77.447&0.93&$-$0.90&50.0$\pm$9.4&$-$30.4$\pm$5.4 \\
78.373&$-$0.48&$-$0.98&36.7$\pm$3.4&$-$49.8$\pm$2.7 \\
79.320&$-$0.58&$-$0.92&33.1$\pm$5.2&$-$53.9$\pm$4.5 \\
\enddata
\tablecomments{Columns are as follows: (1) - Julian date minus 2454000;
(2) - correlation coefficient between Stokes parameters $Q$ and $I$;
(3) - correlation coefficient between Stokes parameters $U$ and $I$;
(4) - degree of polarization; (5) - position angle of polarization.}
\end{deluxetable}

\clearpage

\clearpage

\end{document}